\documentclass[a4paper,fleqn,usenatbib]{mnras}

\usepackage[T1]{fontenc}
\usepackage{ae,aecompl}

\usepackage{graphicx}	
\usepackage{amsmath}	
\usepackage{amssymb}	
\usepackage{color}

\definecolor{grey}{rgb}{0.7,0.7,0.7}

\newcommand{\bpass}{\texttt{bpass}}
\newcommand{\webb}{{\em Webb}}

\title[Recalibrating the Cosmic Star Formation History]{Recalibrating the Cosmic Star Formation History}

\author[Stephen M. Wilkins et al.]{
Stephen M. Wilkins,$^{1}$\thanks{E-mail: s.wilkins@sussex.ac.uk}
Christopher C. Lovell,$^{1,2}$ 
Elizabeth R. Stanway,$^{3}$
\\
$^1$\,Astronomy Centre, Department of Physics and Astronomy, University of Sussex, Brighton, BN1 9QH, UK \\
$^2$\,Centre for Astrophysics Research, School of Physics, Astronomy \& Mathematics, University of Hertfordshire, Hatfield, AL10 9AB, UK \\
$^3$\,Department of Physics, University of Warwick, Gibbet Hill Road, Coventry, CV4 7AL, UK\\
}

\date{Accepted XXX. Received YYY; in original form ZZZ}

\pubyear{2019}

\begin{document}
\label{firstpage}
\pagerange{\pageref{firstpage}--\pageref{lastpage}}
\maketitle

\begin{abstract}

The calibrations linking observed luminosities to the star formation rate depend on the assumed stellar population synthesis model, initial mass function, star formation and metal enrichment history, and whether reprocessing by dust and gas is included. Consequently the shape and normalisation of the inferred cosmic star formation history is sensitive to these assumptions. Using v2.2.1 of the Binary Population and Spectral Synthesis (\bpass) model we determine a new set of calibration coefficients for the ultraviolet, thermal-infrared, and, hydrogen recombination lines. These ultraviolet and thermal infrared coefficients are 0.15-0.2 dex higher than those widely utilised in the literature while the H$\alpha$ coefficient is $\sim 0.35$ dex larger. These differences arise in part due to the inclusion binary evolution pathways but predominantly reflect an extension in the IMF to 300 $M_{\odot}$ and a change in the choice of reference metallicity. We use these new coefficients to recalibrate the cosmic star formation history, and find improved agreement between the integrated cosmic star formation history and the in-situ measured stellar mass density as a function of redshift. However, these coefficients produce new tension between star formation rate densities inferred from the ultraviolet and thermal-infrared and those from H$\alpha$.

\end{abstract}


\section{Introduction}

The recent star formation rate (SFR) is a critical intrinsic parameter of a galaxy which is of crucial importance to understanding the build up of stellar mass in the Universe throughout its history.
To infer the SFR of an observed galaxy from limited wavelength coverage of individual observations of its integrated light a number of calibrations have been determined in the literature, utilising various regions of the rest-frame spectrum \citep{kennicutt_star_1998,murphy_calibrating_2011,kennicutt_star_2012,calzetti_star_2012}.
Emission at UV and optical wavelengths directly probes the light escaping from the photospheres of young, massive stars, however dust extinction complicates this relation, particularly at cosmic noon ($z \sim 2$) where the majority of star formation, at least in massive galaxies, is dust obscured.

The UV / optical light absorbed by dust is thermally reprocessed and re-emitted in the thermal-infrared (TIR, $3-1000\mu$m). In the absence of other sources of heating  \citep[e.g. AGN or older stellar populations][]{lonsdale_persson_origin_1987} the integrated TIR luminosity then traces intrinsic UV and optical emission and thus ongoing star formation. Moreover, deriving integrated TIR luminosities from single band detections introduces uncertainties due to the need to assume a thermal reemission spectrum.

Despite these drawbacks, such calibrations are useful due to their simplicity, allowing them to be applied to large catalogues with ease \citep[e.g. ][]{brinchmann_physical_2004,daddi_multiwavelength_2007,salim_uv_2007}.
However, any biases can impact the derived SFR and any subsequent science interpretations or pipelines reliant on these parameters.

Typically such calibrations are derived using the results of Stellar Population Synthesis (SPS) models, that predict the emission from a simple stellar population given some choice of intrinsic parameters, including the star-formation and enrichment history \citep{conroy_modeling_2013}.
However, there has been significant recent progress in modern SPS models, incorporating a number of physical effects of importance throughout the UV to Far-IR range, such as Thermally Pulsating-Asymptotic Giant Branch stars \citep{conroy_propagation_2009,conroy_propagation_2010}, rotation, nebular emission \citep{byler_nebular_2017} and binary systems \citep{eldridge_binary_2017}.
The majority of stars in the Universe are in binary or higher multiple populations, and this multiplicity fraction is highly mass dependent.
For massive ($>8\,{\rm M_{\odot}}$) and very massive ($>100\,{\rm M_{\odot}}$) stars the multiple fraction is consistent with unity and the mean number of companions actually exceeds one \citep[see recent summary by][]{moe_mind_2017}.
The interaction of a star with its binary companion can have a significant impact on both its emission and any subsequent evolution; for example, contact binaries have much higher surface temperatures which lead to harder UV emission.
Stars in binaries may also be spun up by episodic mass transfer, and evolve as rejuvenated or hot, rotationally mixed stars.
Many interactions of this kind will extend the lifetime of a massive star, leading to stellar populations which remain UV luminous far beyond the few tens of Myr expected for isolated massive stars \citep[see e.g.][]{stanway_interpreting_2014}.

The Binary Population and Spectral Synthesis (\bpass) project \citep{eldridge_binary_2017, stanway_re-evaluating_2018}, incorporates a range of binary interaction effects and their consequences into both a custom grid of stellar evolution models, and a set of stellar population synthesis models derived from these.
Binary interactions considered include Roche lobe overflow and mass transfer, stripping, common envelope evolution phases, simple rotational mixing and rejuvenation due to angular momentum transfer.
Importantly, \bpass\ also includes evolution tracks for very massive stars (M$_{\mathrm{ZAMS}}>100$\,M$_{\odot}$) motivated by observed examples identified by the VLT-FLAMES Tarantula Survey \citep[e.g][]{bestenlehner_vlt-flames_2014}.
Each population is generated with a defined stellar initial mass function (IMF, currently selected from one of nine options) and binary population parameters \citep{moe_mind_2017}.
In addition to data products including stellar population and transient data, they yield an integrated light spectral energy distribution for simple stellar populations as a function of age at 13 metallicities. These can be combined with a star formation history to create complex stellar populations, and processed with radiative transfer models to account for the effects of nebular gas, before being convolved with filter profiles to simulate photometric observations. The current version of \bpass\ is v2.2.1 \citep{stanway_re-evaluating_2018} and we use this to explore the effect of binary populations on SFR calibrations.

The Cosmic Star Formation Rate Density (CSFRD) is a key diagnostic of stellar assembly in the Universe.
It can be measured through direct, point-in-time observations of SFRs of galaxies at different redshifts \citep[e.g.][]{madau_high-redshift_1996, lilly_canada-france_1996, hopkins_normalization_2006, madau_cosmic_2014}, or through archaeological approaches that estimate the star formation history of local galaxies \citep[e.g.][]{panter_star_2007, tojeiro_recovering_2007}.
Both approaches produce similar predictions for the shape of the CSFRD; it rises at early times to a peak at cosmic noon ($z \sim 2$), then falls by approximately an order of magnitude to the present day \citep{madau_cosmic_2014}.
The CSFRD is closely linked to the evolution of the cosmic stellar mass density, which is simply the integral of the CSFRD over time, assuming some age-dependent recycling factor due to stellar evolution.
\cite{wilkins_evolution_2008}, however, first showed how these two key diagnostics are in tension; at early times, the star formation rate would require a much higher stellar mass density at $z \sim 3$.
A number of explanations for this discrepancy have been suggested in the literature, such as incorrect stellar mass measurements, dust corrections, a variable IMF, or incorrect inferred metallicities \citep{yu_inconsistency_2016}.
The inferred CSFRD is also in tension with that produced by recent numerical simulations, which tend to underestimate the normalisation at cosmic noon \citep{dave_galaxy_2008,furlong_evolution_2015,diemer_log-normal_2017}.
We here explore the impact on this discrepancy of updated SFR calibrations.

This study is arranged as follows.
In Section \ref{sec:c} we show how using the \bpass\ models leads to changes in the SFR calibration coefficients derived by \cite{murphy_calibrating_2011} and promoted by \cite{kennicutt_star_2012}.
In Section \ref{sec:csfh} we show the impact of these newly derived calibrations on the evolution of the CSFRD, and show the knock-on effect on the discrepancy with the cosmic stellar mass density evolution.
Finally, in Section \ref{sec:conc} we discuss our results and state our conclusions.

\section{Calibrations}\label{sec:c}

We follow \citet{kennicutt_star_2012} and define SFR calibration coefficients as
\begin{equation}
C_{x} = \frac{L_x \,/\, {\rm erg \; s^{-1}}}{{\rm SFR} \,/\, {\rm M_{\sun}\, yr^{-1}}} \,\,,
\end{equation}
where $x$ is the band or line chosen.
By default we assume v2.2.1 of \bpass\footnote{\href{https://bpass.auckland.ac.nz}{https://bpass.auckland.ac.nz}}, adopt a \citet{chabrier_galactic_2003} IMF from $0.1-300\,{\rm M_{\sun}}$, and include the effects of nebular continuum and line emission using the 2017 version \citep{ferland_2017_2017} of the \texttt{cloudy} photo-ionisation model\footnote{\href{https://www.nublado.org/}{https://www.nublado.org/}}. The widely used set of coefficients derived by \citet{murphy_calibrating_2011} assumed the \texttt{Starburst99} SPS model \citep{leitherer_starburst99:_1999}, a \citet{kroupa_variation_2001} IMF (a slope of 1.3 between 0.1 and 0.5 ${\rm M_{\odot}}$ and 2.3 between 0.5 and 100 ${\rm M_{\odot}}$), included nebular continuum emission, with coefficients evaluated assuming 100 Myr previous constant star formation and $Z=0.02$.

\subsection{Far UV}

\begin{figure}
\centering
\includegraphics[width=20pc]{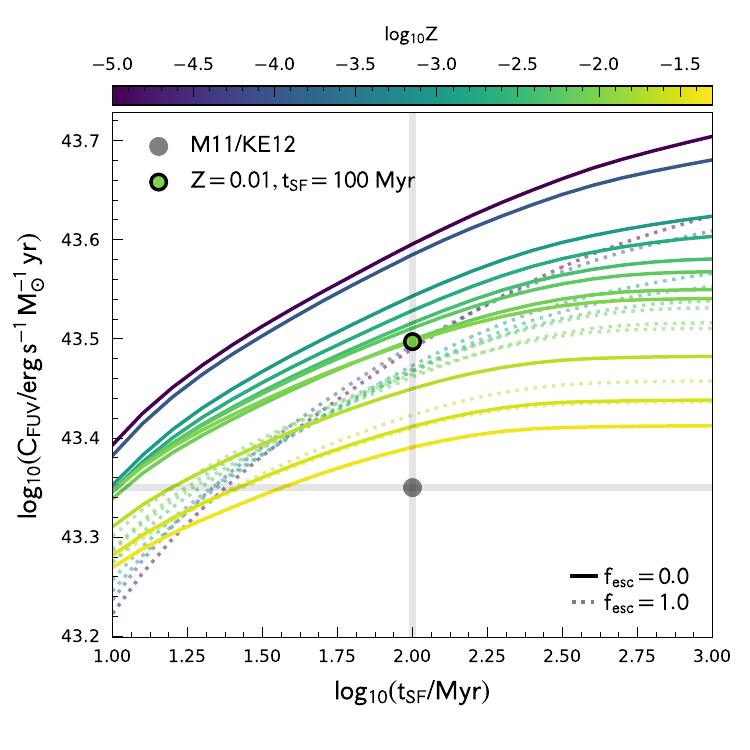}
\caption{Sensitivity of the FUV calibration to the duration of previous (constant) star formation and metallicity assuming our default model (BPASSv2.2.1, Chabrier IMF, $m_{up}=300\, {\rm M_{\odot}}$.). Dotted lines show results assuming no nebular continuum or line contribution. The coloured point denotes our default calibration assuming 100 Myr previous star formation and $Z=0.01$. The grey point denotes the \citet{murphy_calibrating_2011} calibration.}
\label{fig:UV}
\end{figure}

We begin, in Figure \ref{fig:UV} by showing the sensitivity of the far-UV (FUV, $\lambda/\mu\text{m}\in [0.14, 0.16]$) calibration coefficient to the duration of previous star formation, metallicity, and the inclusion of nebular emission.
In summary, the greater the duration of the star formation episode the higher the calibration, though it begins to plateau at $\sim$ 250 Myr, where the UV emission from older populations is reduced through stellar evolution and recycling.
Lower metallicities also lead to a higher coefficient, due to the increased hardness of the UV emission in these populations.
The inclusion of nebular emission changes the shape and normalisation of the relation, but the trends from the pure stellar case are mostly preserved.

\begin{figure}
\centering
\includegraphics[width=20pc]{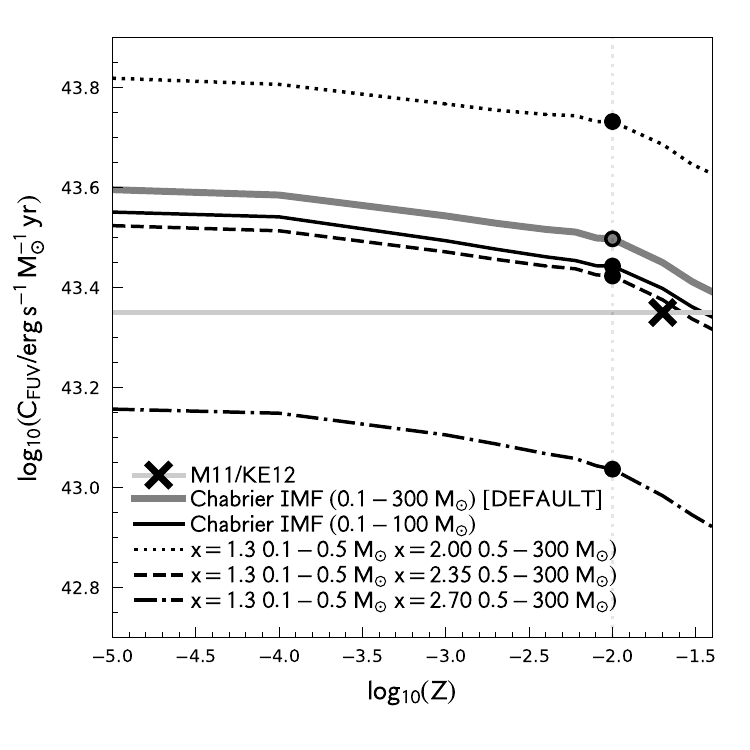}
\caption{Sensitivity of the FUV calibration to the choice of SPS model, IMF, and metallicity. In this case we assume constant previous star formation of 100 Myr.}
\label{fig:UV_SPSIMF}
\end{figure}

In Figure \ref{fig:UV_SPSIMF} we also explore the impact of adopting an alternative IMF on the calibration, this time focussing on the trend with metallicity by fixing the duration of previous star formation at $100\,{\rm Myr}$ and $f_{\rm esc}=0.0$. Reducing the high-mass cut off of the IMF from $300$ to $100\,{\rm M_{\odot}}$ yields calibrations consistently around 0.05 dex lower. Adopting a broken power law form of the IMF with a Salpeter-like ($-2.35$) high-mass ($>0.5\,{\rm M_{\odot}}$) slope and a flatter ($-1.3$) low-mass slope yields calibrations lower by $\approx 0.07$ dex. We also show more dramatic changes to the high-mass slope: adopting a shallow ($-2.0$) slope yields a calibration $\approx 0.2$ dex higher while a very steep ($-2.7$) slope reduces the calibration by $\approx 0.4$ dex.

For our fiducial calibration we assume 100 Myr previous continuous star formation and $Z=0.01$. The resulting coefficient is $\log_{10}(C_{\rm FUV})=43.50$ \citep[c.f. $43.35$ assumed in][]{murphy_calibrating_2011}. This increase is a consequence of four factors: the extension of the IMF to 300 ${\rm M_{\odot}}$ ($+0.06$ dex); the use of $Z=0.01$ as the reference metallicity instead of $Z=0.02$ ($+0.04$ dex); the adoption of a \citet{chabrier_galactic_2003} IMF instead of a \citet{kroupa_variation_2001} ($+0.025$ dex); and differences in the SPS model ($+0.025$ dex).

\subsection{Thermal IR}

\begin{figure}
\centering
\includegraphics[width=20pc]{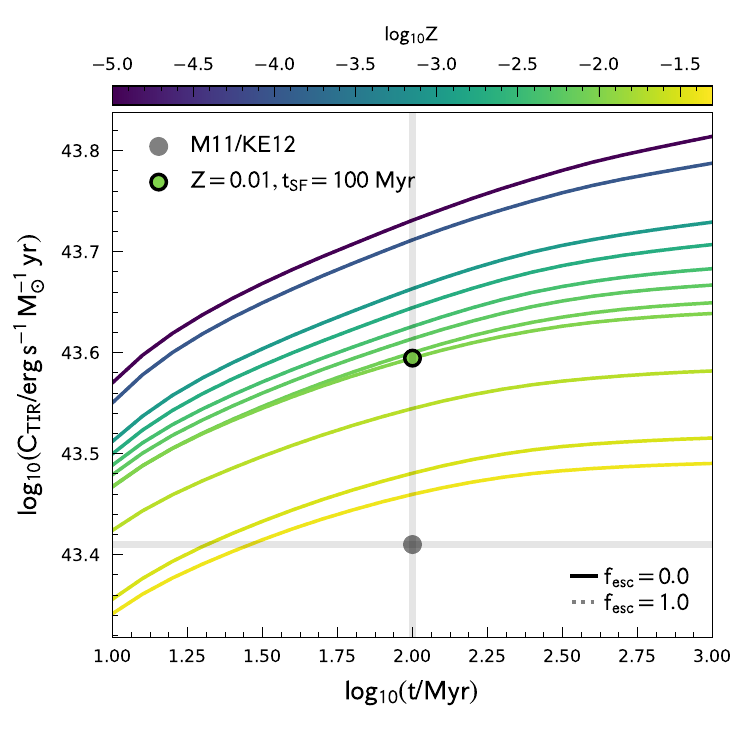}
\caption{As for Figure \ref{fig:UV}, but showing the thermal infrared calibration ($C_\mathrm{TIR}$).}
\label{fig:TIR}
\end{figure}

\begin{figure}
\centering
\includegraphics[width=20pc]{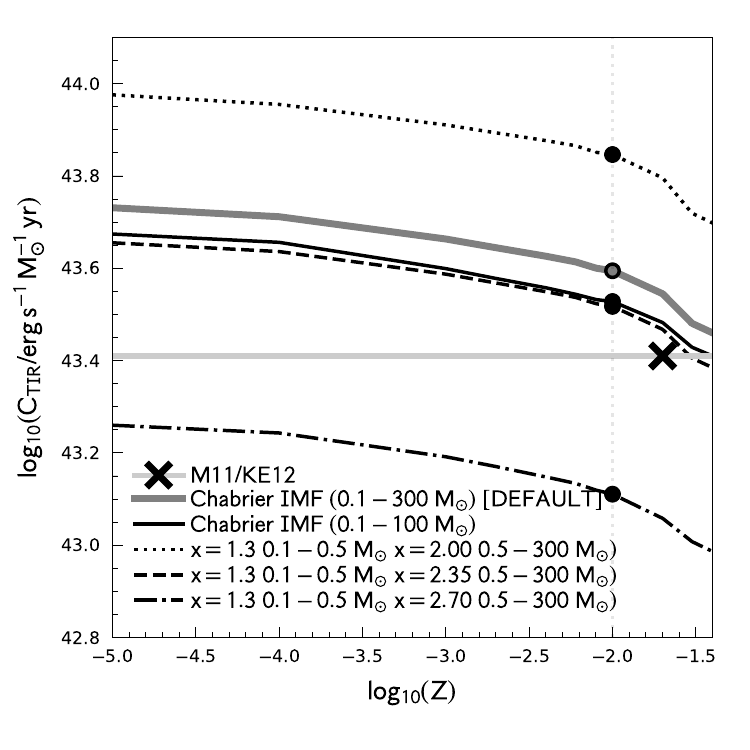}
\caption{The same a \ref{fig:UV_SPSIMF} but for the thermal infrared calibration.}
\label{fig:TIR_SPSIMF}
\end{figure}

In most galaxies some fraction of the UV/optical light is absorbed by dust and thermally reprocessed into the IR. To define the TIR calibration we follow \citet{murphy_calibrating_2011} and assume the entire Balmer continuum is absorbed by dust and re-emitted in the TIR. The sensitivity of this calibration to the assumed duration of previous star formation, metallicity, and the inclusion of nebular emission is shown in Figure \ref{fig:TIR}. For our default set of assumptions we obtain $\log_{10}(C_{\rm TIR})=43.59$ \citep[c.f. $43.41$ obtained by][]{murphy_calibrating_2011}. Figure \ref{fig:TIR_SPSIMF} shows the sensitivity of $C_{\rm TIR}$ to the choice of IMF. Unsurprisingly, given the overlapping wavelength range, the TIR calibration follows broadly the same trends as the far-UV calibration. If instead we assume that all stellar (and nebular) emission is reprocessed $C_{\rm TIR}$ would increase by $\approx 0.1$ dex.

\subsection{Hydrogen Recombination Lines}

In galaxies dominated by stellar emission virtually all the H ionising (Lyman continuum or LyC) photons are produced by young, massive stars.
As these stars, due to their short lifetimes, are contemporaneous with star formation events the LyC production rate is an ideal diagnostic of star formation activity.
While some fraction of LyC may escape, the vast majority are reprocessed into recombination lines.
Unlike most metal lines, the link between the LyC production rate and the H recombination line luminosity is not strongly dependent on the metallicity, temperature, or density of the reprocessing gas.

While the brightest H recombination line, Lyman-$\alpha$, is unsuitable due to resonant scattering and susceptibility to dust, the second brightest line, H$\alpha$ is ideally located in the optical, making it a favoured star formation rate diagnostic, particularly at low/intermediate redshift where it remains accessible to ground based observatories.
While presently of limited use at $z>3$ the advent of deep $\lambda>2\mu$m near-IR spectroscopy obtainable by the {\em Webb Telescope} will enable the wide use of H$\alpha$ to $z\approx 6$ and potentially beyond.

\begin{figure}
\centering
\includegraphics[width=20pc]{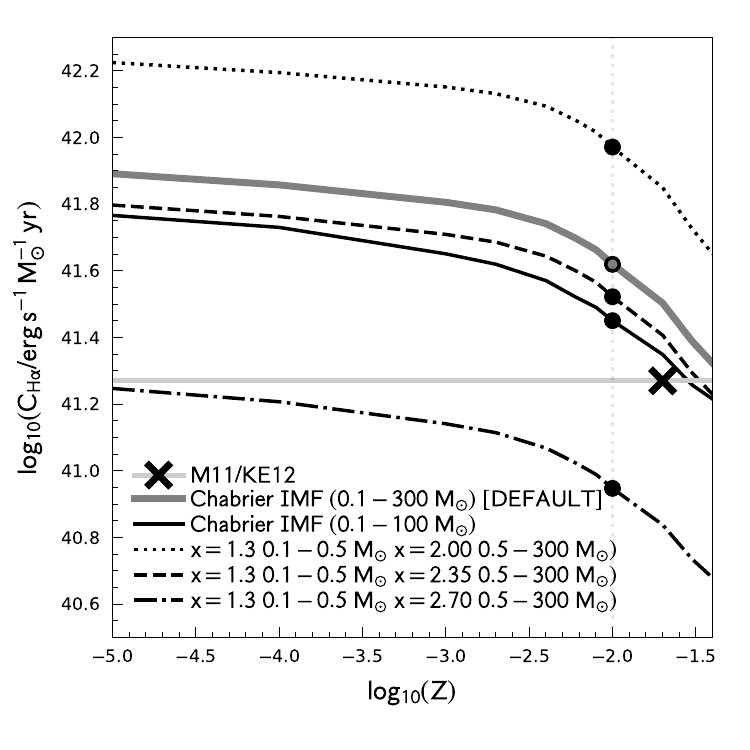}
\caption{Sensitivity of the H$\alpha$ calibration to the choice of SPS model, IMF, and metallicity. Assumes constant previous star formation of 100 Myr.}
\label{fig:Ha}
\end{figure}

To calculate the H$\alpha$ calibration we calculate the LyC production rate and assume Case B recombination assuming $n_{e}=10^{2}\,{\rm cm^{-3}}$ and $T=10^{4}\,{\rm K}$ \citep{osterbrock_astrophysics_2006}.
As the LyC production rate is dominated by the most massive stars for constant star formation there is very little trend with duration.
In Figure \ref{fig:Ha} we thus focus on the sensitivity of $C_{\rm H \alpha}$ to the metallicity for a range of IMFs. The resulting calibration for our default assumptions, $\log_{10}(C_{\rm H\alpha})=41.62$, is $0.35$ dex larger than that derived by \citet{murphy_calibrating_2011}. In this case $\approx 0.15$ dex of this difference is now attributed to the extension of the IMF to $300\, {\rm M_{\sun}}$ while $\approx 0.12$ dex is due to the difference refence metallicity and 0.025 dex is due to the difference between the \citet{chabrier_galactic_2003} and \citet{kroupa_variation_2001} IMF. Thus, $\approx 0.075$ dex is attributable to differences between \bpass\ and \texttt{Starburst99}.

Due to the fact that LyC emission is dominated by more massive stars $C_{\rm H\alpha}$ shows a stronger trend with the high-mass slope of the IMF. Assuming a steep ($-2.7$) high-mass slope yields a calibration approximately $0.65$ dex smaller than based on our default assumptions.

\webb's ability to obtain deep near-IR spectroscopy at $>2\mu$m also opens the possibility of using the Paschen-$\alpha$ line ($\lambda=1.875\mu$m) to $z\sim 1.5$. This is advantageous as Paschen-$\alpha$ will be much less susceptible to dust than H$\alpha$. Assuming the same Case B assumptions as above $\log_{10}(C_{\rm Pa-\alpha}) = \log_{10}(C_{\rm H\alpha})-0.93 = 40.69$.

\subsection{Discussion}

In this section we have calculated the SFR calibration coefficients for the far-UV, thermal-IR, and H$\alpha$ assuming the \bpass\ stellar population synthesis model. In each case we obtain calibration coefficients that are between $0.15-0.35$ dex higher than adopted by \citet{murphy_calibrating_2011} \citep[see also][]{kennicutt_star_2012}. These calibrations, and their analogues in \citet{murphy_calibrating_2011}, are stated again in Table \ref{tab:C_parameters} for completeness.

The higher calibration coefficients found by this study result from both the choice of an alternative IMF and reference metallicity but also the addition of very massive ($>100\,{\rm M_{\odot}}$) stars and the hotter ultraviolet spectra of stars resulting from binary interactions. Both are physically motivated by observations in the local universe, but it should be noted that uncertainties remain in the extent and sampling of the very massive star initial mass function in a typical galaxy, the far-ultraviolet rest frame emission spectrum of the hottest binary products, and their extrapolations to very low metallicities.

\begin{table}
 \caption{Fiducial calibration coefficients from \bpass\ assuming a range of assumptions, and \citet{murphy_calibrating_2011} (M11). A: \citet{murphy_calibrating_2011} coefficient. Assumes \citet{kroupa_variation_2001} IMF and $m_{up}=100\, {\rm M_{\odot}}$ and $Z=0.02$ reference metallicity. B: as A but assuming \bpass. C: as B but assuming the \citet{chabrier_galactic_2003} IMF. D: as C but assuming $m_{up}=300\, {\rm M_{\odot}}$. E: as D but assuming $Z=0.01$ as the reference metallicity. This is our default set of assumptions.
}
 \label{tab:C_parameters}
 \begin{tabular}{lcccccc}
  \hline
    & \multicolumn{5}{c}{$\log_{10}(C)$}  \\
    \hline
    & M11 & \multicolumn{4}{c}{\bpass} \\
    \hline
    & A & B & C & D & {\bf E} \\
  \hline
UV         & 43.35 & 43.38 & 43.40 & 43.46 & {\bf 43.50}  \\
TIR        & 43.41 & 43.45 & 43.48 & 43.54 & {\bf 43.59}  \\
H$\alpha$  & 41.27 & 41.35 & 41.37 & 41.52 & {\bf 41.62}  \\
Pa$\alpha$ & -     & 40.42 & 40.44 & 40.59 & {\bf 40.69}  \\
  \hline
  \hline
 \end{tabular}
\end{table}

\section{Cosmic Star Formation History}\label{sec:csfh}

As we have shown, our new calibrations are between $0.15-0.35$ dex larger than those routinely utilised in the literature. We now investigate the impact of assuming these calibrations on the cosmic star formation history (CSFH) and the growth of stellar mass.

\subsection{Ultraviolet and Thermal Infrared}

\begin{figure}
\centering
\includegraphics[width=20pc]{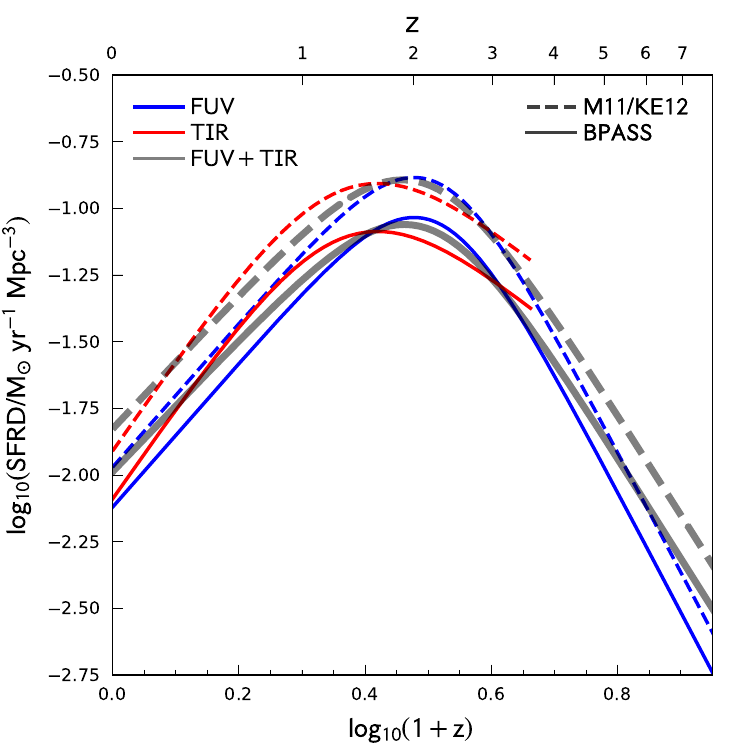}
\caption{The \citet{madau_cosmic_2014} star formation rate density (SFRD) compilation recalibrated to assume both the \citet{murphy_calibrating_2011} calibrations and calibrations based on \bpass. The solid curves show best fits to the individual SFRD measurements using the parameterisation employed by \citet{madau_cosmic_2014}. }
\label{fig:CSFH_UVTIR}
\end{figure}

We begin by focussing on the CSFH as probed by the UV and TIR. To do this we make use of the relatively recent compilation of luminosity densities (and dust corrections) by \citet{madau_cosmic_2014}. We re-calibrate\footnote{\citet{madau_cosmic_2014} assumed an independent set of calibrations.} this compilation to assume both the \citet{murphy_calibrating_2011} calibrations and our new calibrations. We then fit UV, TIR, and combined compilation by the same simple parameterisation of the CSFH utilised by \citet{madau_cosmic_2014}:
\begin{equation}
\psi = \frac{a(1+z)^{b}}{1+\left(\left(1+z\right)/c\right)^{d}}
\end{equation}
The resulting best-fit parameters are provided in Tab. \ref{tab:CSFH_parameters} and the CSFHs are shown in Fig. \ref{fig:CSFH_UVTIR}. The main difference is the change in normalisation $a$. Adopting our \bpass\ based calibrations results in a normalisation that is 0.16 dex (30\%) lower.

\begin{table}
 \caption{Best fit CSFH parameters. $^{1}$ units of ${\rm M_{\sun}\, yr^{-1}\, Mpc^{-3}}$.}
 \label{tab:CSFH_parameters}
 \begin{tabular}{lcccc}
  \hline
    & $a^{1}$ & $b$ & $c$ & $d$\\
  \hline
  & \multicolumn{4}{|c|}{\citet{murphy_calibrating_2011}}\\
UV & 0.0107 & 2.70 & 3.22 & 7.22 \\
TIR & 0.0124 & 3.43 & 2.37 & 5.38 \\
UV+TIR  & 0.0149 & 2.52 & 3.07 & 6.27 \\
  \hline
 & \multicolumn{4}{|c|}{\bpass}\\
UV & 0.0087 & 2.70 & 3.22 & 7.22 \\
TIR & 0.0082 & 3.43 & 2.37 & 5.38 \\
UV+TIR & 0.0103 & 2.48 & 3.10 & 6.26 \\
  \hline
 \end{tabular}
\end{table}

\subsection{H$\alpha$}

\begin{figure}
\centering
\includegraphics[width=20pc]{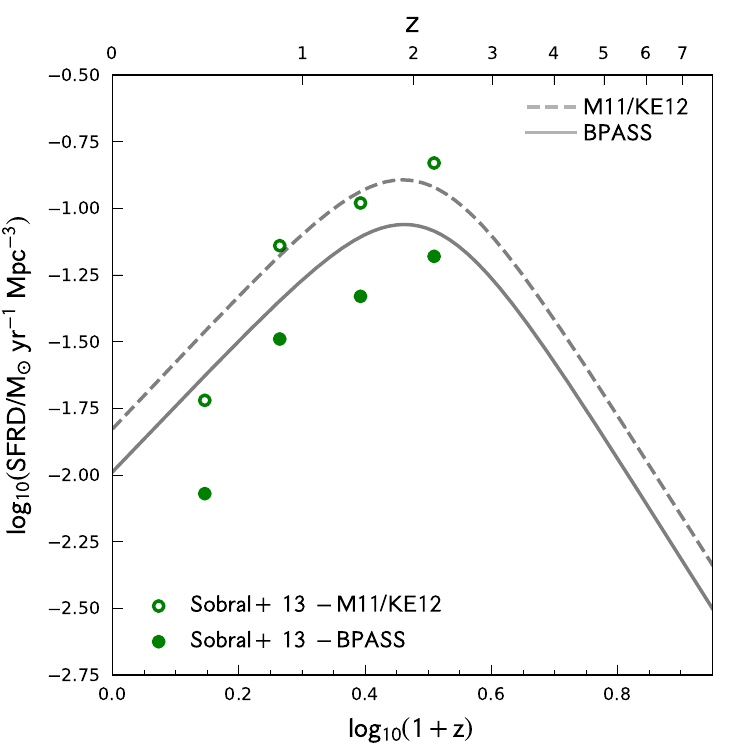}
\caption{Our UV+TIR CSFH fits assuming the \citet{murphy_calibrating_2011} (M11) and \bpass\ calibrations compared with H$\alpha$ inferred star formation rate densities from \citet{sobral_large_2013} again assuming the M11 and \bpass\ calibrations.}
\label{fig:CSFH_Ha}
\end{figure}

While currently not as ubiquitous as the use of UV and TIR (particularly at high-redshift), various studies have measured the H$\alpha$ luminosity function either via narrow-band imaging surveys \citep[e.g.][]{geach_hizels:_2008, dale_wyoming_2010, ly_hensuremathalpha_2011, sobral_large_2013}, targeted spectroscopic campaigns \citep[e.g.][]{gunawardhana_galaxy_2013}, or using wide field slitless spectroscopy. \citet{sobral_large_2013} measured H$\alpha$ inferred CSFRD, and found good agreement with those presented in \citet{murphy_calibrating_2011}.
To test whether this is still the case, we show in Fig. \ref{fig:CSFH_Ha} the UV+TIR CSFRD and the H$\alpha$ CSFRD assuming both the \citet{murphy_calibrating_2011} calibrations and our \bpass\ calibrations.

While there is good agreement between the two CSFRDs at cosmic noon assuming \citet{murphy_calibrating_2011}, this is not true for our new \bpass\ based calibration coefficients with H$\alpha$ inferred star formation rate densities lying 0.1-0.2 dex below those based on the UV or TIR. There are several possible solutions: our modelling changes are incorrect; the high-mass slope of the IMF could be steeper; or, observations of the H$\alpha$ LF are incomplete, perhaps missing a contribution from heavily obscured systems.

With its state-of-the-art near-IR spectroscopic capabilities, the {\em Webb Telescope} should provide a more definitive answer as to whether this discrepancy remains. This is possible not only with the H$\alpha$ line but also Paschen-$\alpha$ which is less suceptible to the effects of dust attenuation.

\subsection{Assembly of Stellar Mass}

The stellar mass density ($\rho_{\star}$) is related to the integral of the past star formation history (${\rm CSFRD}(t)$) through
\begin{equation}
\rho_{\star} = \int \left[1 - f_{r}(t)\right]{\rm CSFRD}(t)\ {\rm d}t \,\,,
\end{equation}
where $t$ is defined as the age and $f_{r}$ is the fraction of material returned to the inter-stellar medium. $f_{r}$ is a strong function of age with a weaker dependence on metallicity. $f_{r}$ is also sensitive to details of the SPS model.

\begin{figure}
\centering
\includegraphics[width=20pc]{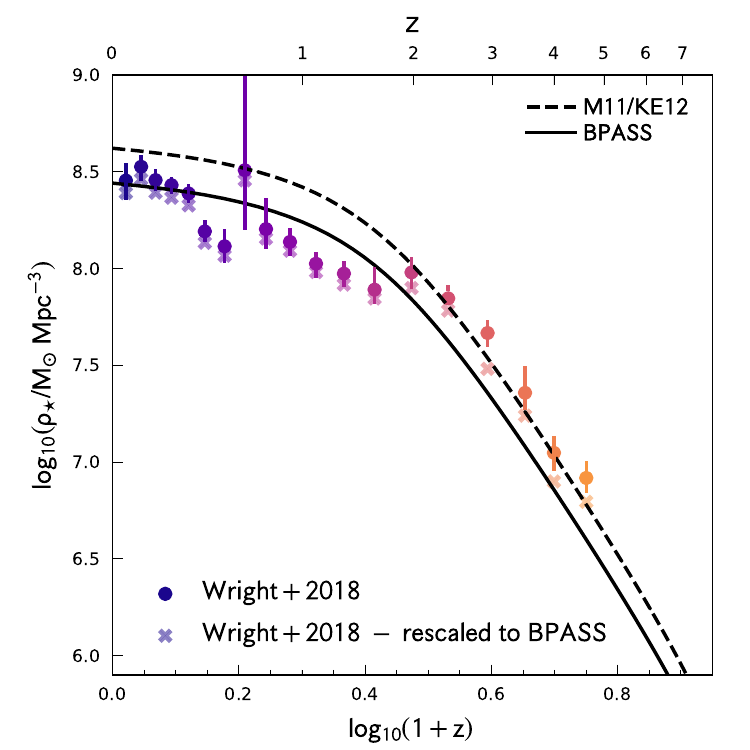}
\caption{The stellar mass density of the Universe. The two lines show the stellar mass density inferred from the integral of the cosmic star formation history (accounting for the return of material to the ISM). Circles show the recent multi-field estimates from \citet{wright_gama/g10-cosmos/3d-hst:_2018}. The lower set of points (crosses) show the inferred stellar mass densities from \citet{wright_gama/g10-cosmos/3d-hst:_2018} but with an approximate correction to assume \bpass.}
\label{fig:SMD}
\end{figure}

The assembly of stellar mass density inferred from the CSFRD, assuming both the \citet{murphy_calibrating_2011} calibration coefficients and our \bpass\ inferred coefficients, is shown in Fig. \ref{fig:SMD}. At $z=0$ this yields $\log_{10}(\rho_{\star}/{\rm M_{\sun}\, Mpc^{-3}})\approx 8.62$ and $8.44$ assuming the \citet{murphy_calibrating_2011} and our coefficients, respectively.

Fig. \ref{fig:SMD} also shows recent observational constraints on the stellar mass density from \cite{wright_gama/g10-cosmos/3d-hst:_2018}
who utilised observations from GAMA \citep{driver_gama:_2009,driver_galaxy_2011,driver_gama/g10-cosmos/3d-hst:_2018},
COSMOS \citep{andrews_g10/cosmos:_2017},
and 3D-HST \citep{brammer_3d-hst:_2012,momcheva_3d-hst_2016,skelton_3d-hst_2014} to measure galaxy stellar masses, and thus stellar mass densities, from $z=0-5$.
At $z<2$ these in-situ estimates all lie below the prediction assuming the \citet{murphy_calibrating_2011} calibration coefficients. However, such a comparison is not self-consistent as \cite{wright_gama/g10-cosmos/3d-hst:_2018} utilise \cite{bruzual_stellar_2003} models to infer their stellar mass estimates.
To be completely self-consistent all individual stellar masses should be remeasured using the \bpass\ models.
A rough estimate of the impact of instead assuming \bpass\ can, however, be made by comparing the optical ($V$-band) mass-to-light ratio (M/L) predicted by BC03 and \bpass\ for the same star formation history.
Using the \bpass\ inferred CSFRD we calculate the $V$-band mass-to-light ratio assuming both \bpass\ and BC03 and use the offset to rescale the \cite{wright_gama/g10-cosmos/3d-hst:_2018} stellar mass densities.
At $z=0$ the magnitude of this adjustment is $\approx 0.06$ dex, decreases to $\approx 0.04$ at $z\approx 1$, before increasing to $0.1-0.2$ dex at $z>2.5$.
These ``rescaled" stellar mass densities are shown by crosses in Fig. \ref{fig:SMD}. At $z<0.5$ and $z>2$ these are in good agreement with the \bpass\ predictions. At $0.5<z<2$ the overall agreement is less good, though nevertheless better than assuming the \citet{murphy_calibrating_2011} coefficients. At $z>2$ observed stellar mass densities are larger than predicted assuming our new coefficients with \citet{murphy_calibrating_2011} coefficients providing a better fit.

\section{Conclusions}\label{sec:conc}

Using the \bpass\ stellar population synthesis model we have derived new calibrations relating star formation to the far-UV, thermal-IR, and H$\alpha$ luminosity.
These calibrations coefficients are $0.15-0.35$ dex larger than those derived by \citet{murphy_calibrating_2011} and promoted by \citet{kennicutt_star_2012}.
These differences arise due several different choices: 1) the assumption of a different IMF, in particular the extension to $300\, {\rm M_{\sun}}$; 2) differences between stellar population synthesis model and specifically the inclusion of binary pathways; and 3) the choice of $Z=0.01$ as our fiducial metallicity.
The difference is largest for the H$\alpha$ calibration ($0.35$ dex); this is particularly important as both H$\alpha$ and other hydrogen recombination lines (in particular Paschen-$\alpha$) will gain increased prominence as tracers of star formation at high-redshift with the arrival of the {\em Webb Telescope}.

We then recalibrate the \citet{madau_cosmic_2014} star formation rate density compilation, obtaining a cosmic star formation rate density (CSFRD) $\approx 0.2$ dex (35\%) lower than assuming the \citet{murphy_calibrating_2011} calibrations coefficients.
This CSFRD provides better agreement with recent observational constraints on the in-situ assembly of stellar mass, coming close to reconciling the tension first proposed in \cite{wilkins_extragalactic_2008}.
However, the updated CSFRD relations from  far-UV and thermal-IR tracers are in tension with those inferred from H$\alpha$ which are now systematically $0.1-0.2$ dex too low raising concerns of an outstanding issue with either the calibrations or the observations.

\subsection*{Acknowledgements}

CCL acknowledges support from an STFC studentship. ERS acknowledges support from STFC consolidated grant ST/P000495/1.

\bibliographystyle{mnras}
\bibliography{SFRCalib}

\begin{thebibliography}{}
\makeatletter
\relax
\def\mn@urlcharsother{\let\do\@makeother \do\$\do\&\do\#\do\^\do\_\do\%\do\~}
\def\mn@doi{\begingroup\mn@urlcharsother \@ifnextchar [ {\mn@doi@}
  {\mn@doi@[]}}
\def\mn@doi@[#1]#2{\def\@tempa{#1}\ifx\@tempa\@empty \href
  {http://dx.doi.org/#2} {doi:#2}\else \href {http://dx.doi.org/#2} {#1}\fi
  \endgroup}
\def\mn@eprint#1#2{\mn@eprint@#1:#2::\@nil}
\def\mn@eprint@arXiv#1{\href {http://arxiv.org/abs/#1} {{\tt arXiv:#1}}}
\def\mn@eprint@dblp#1{\href {http://dblp.uni-trier.de/rec/bibtex/#1.xml}
  {dblp:#1}}
\def\mn@eprint@#1:#2:#3:#4\@nil{\def\@tempa {#1}\def\@tempb {#2}\def\@tempc
  {#3}\ifx \@tempc \@empty \let \@tempc \@tempb \let \@tempb \@tempa \fi \ifx
  \@tempb \@empty \def\@tempb {arXiv}\fi \@ifundefined
  {mn@eprint@\@tempb}{\@tempb:\@tempc}{\expandafter \expandafter \csname
  mn@eprint@\@tempb\endcsname \expandafter{\@tempc}}}

\bibitem[\protect\citeauthoryear{Andrews, Driver, Davies, Kafle, Robotham  \&
  Wright}{Andrews et~al.}{2017}]{andrews_g10/cosmos:_2017}
Andrews S.~K.,  Driver S.~P.,  Davies L. J.~M.,  Kafle P.~R.,  Robotham A.
  S.~G.,   Wright A.~H.,  2017, \mn@doi [Mon Not R Astron Soc]
  {10.1093/mnras/stw2395}, 464, 1569

\bibitem[\protect\citeauthoryear{Bestenlehner et~al.,}{Bestenlehner
  et~al.}{2014}]{bestenlehner_vlt-flames_2014}
Bestenlehner J.~M.,  et~al., 2014, \mn@doi [{\textbackslash}aap]
  {10.1051/0004-6361/201423643}, 570, A38

\bibitem[\protect\citeauthoryear{Brammer et~al.,}{Brammer
  et~al.}{2012}]{brammer_3d-hst:_2012}
Brammer G.~B.,  et~al., 2012, \mn@doi [ApJS] {10.1088/0067-0049/200/2/13}, 200,
  13

\bibitem[\protect\citeauthoryear{Brinchmann, Charlot, White, Tremonti,
  Kauffmann, Heckman  \& Brinkmann}{Brinchmann
  et~al.}{2004}]{brinchmann_physical_2004}
Brinchmann J.,  Charlot S.,  White S. D.~M.,  Tremonti C.,  Kauffmann G.,
  Heckman T.,   Brinkmann J.,  2004, \mn@doi [Monthly Notices of the Royal
  Astronomical Society] {10.1111/j.1365-2966.2004.07881.x}, 351, 1151

\bibitem[\protect\citeauthoryear{Bruzual \& Charlot}{Bruzual \&
  Charlot}{2003}]{bruzual_stellar_2003}
Bruzual G.,  Charlot S.,  2003, \mn@doi [{\textbackslash}mnras]
  {10.1046/j.1365-8711.2003.06897.x}, 344, 1000

\bibitem[\protect\citeauthoryear{Byler, Dalcanton, Conroy  \& Johnson}{Byler
  et~al.}{2017}]{byler_nebular_2017}
Byler N.,  Dalcanton J.~J.,  Conroy C.,   Johnson B.~D.,  2017, \mn@doi [The
  Astrophysical Journal] {10.3847/1538-4357/aa6c66}, 840, 44

\bibitem[\protect\citeauthoryear{Calzetti}{Calzetti}{2012}]{calzetti_star_2012}
Calzetti D.,  2012, arXiv:1208.2997 [astro-ph]

\bibitem[\protect\citeauthoryear{Chabrier}{Chabrier}{2003}]{chabrier_galactic_2003}
Chabrier G.,  2003, \mn@doi [{\textbackslash}pasp] {10.1086/376392}, 115, 763

\bibitem[\protect\citeauthoryear{Conroy}{Conroy}{2013}]{conroy_modeling_2013}
Conroy C.,  2013, \mn@doi [Annual Review of Astronomy and Astrophysics]
  {10.1146/annurev-astro-082812-141017}, 51, 393

\bibitem[\protect\citeauthoryear{Conroy \& Gunn}{Conroy \&
  Gunn}{2010}]{conroy_propagation_2010}
Conroy C.,  Gunn J.~E.,  2010, \mn@doi [The Astrophysical Journal]
  {10.1088/0004-637X/712/2/833}, 712, 833

\bibitem[\protect\citeauthoryear{Conroy, Gunn  \& White}{Conroy
  et~al.}{2009}]{conroy_propagation_2009}
Conroy C.,  Gunn J.~E.,   White M.,  2009, \mn@doi [The Astrophysical Journal]
  {10.1088/0004-637X/699/1/486}, 699, 486

\bibitem[\protect\citeauthoryear{Daddi et~al.,}{Daddi
  et~al.}{2007}]{daddi_multiwavelength_2007}
Daddi E.,  et~al., 2007, \mn@doi [ApJ] {10.1086/521818}, 670, 156

\bibitem[\protect\citeauthoryear{Dale et~al.,}{Dale
  et~al.}{2010}]{dale_wyoming_2010}
Dale D.~A.,  et~al., 2010, \mn@doi [{\textbackslash}apjl]
  {10.1088/2041-8205/712/2/L189}, 712, L189

\bibitem[\protect\citeauthoryear{Dav{\'e}}{Dav{\'e}}{2008}]{dave_galaxy_2008}
Dav{\'e} R.,  2008, \mn@doi [Mon Not R Astron Soc]
  {10.1111/j.1365-2966.2008.12866.x}, 385, 147

\bibitem[\protect\citeauthoryear{Diemer, Sparre, Abramson  \& Torrey}{Diemer
  et~al.}{2017}]{diemer_log-normal_2017}
Diemer B.,  Sparre M.,  Abramson L.~E.,   Torrey P.,  2017, \mn@doi [The
  Astrophysical Journal] {10.3847/1538-4357/aa68e5}, 839, 26

\bibitem[\protect\citeauthoryear{Driver, Norberg, Baldry, Bamford, Hopkins,
  Liske, Loveday  \& Peacock}{Driver et~al.}{2009}]{driver_gama:_2009}
Driver S.~P.,  Norberg P.,  Baldry I.~K.,  Bamford S.~P.,  Hopkins A.~M.,
  Liske J.,  Loveday J.,   Peacock J.~A.,  2009, \mn@doi [A\&G]
  {10.1111/j.1468-4004.2009.50512.x}, 50, 5.12

\bibitem[\protect\citeauthoryear{Driver et~al.,}{Driver
  et~al.}{2011}]{driver_galaxy_2011}
Driver S.~P.,  et~al., 2011, \mn@doi [Mon Not R Astron Soc]
  {10.1111/j.1365-2966.2010.18188.x}, 413, 971

\bibitem[\protect\citeauthoryear{Driver et~al.,}{Driver
  et~al.}{2018}]{driver_gama/g10-cosmos/3d-hst:_2018}
Driver S.~P.,  et~al., 2018, \mn@doi [Monthly Notices of the Royal Astronomical
  Society] {10.1093/mnras/stx2728}, 475, 2891

\bibitem[\protect\citeauthoryear{Eldridge, Stanway, Xiao, McClelland, Taylor,
  Ng, Greis  \& Bray}{Eldridge et~al.}{2017}]{eldridge_binary_2017}
Eldridge J.~J.,  Stanway E.~R.,  Xiao L.,  McClelland L. A.~S.,  Taylor G.,  Ng
  M.,  Greis S. M.~L.,   Bray J.~C.,  2017, \mn@doi [{\textbackslash}pasa]
  {10.1017/pasa.2017.51}, 34, e058

\bibitem[\protect\citeauthoryear{Ferland et~al.,}{Ferland
  et~al.}{2017}]{ferland_2017_2017}
Ferland G.~J.,  et~al., 2017, {\textbackslash}rmxaa, 53, 385

\bibitem[\protect\citeauthoryear{Furlong et~al.,}{Furlong
  et~al.}{2015}]{furlong_evolution_2015}
Furlong M.,  et~al., 2015, \mn@doi [Monthly Notices of the Royal Astronomical
  Society] {10.1093/mnras/stv852}, 450, 4486

\bibitem[\protect\citeauthoryear{Geach, Smail, Best, Kurk, Casali, Ivison  \&
  Coppin}{Geach et~al.}{2008}]{geach_hizels:_2008}
Geach J.~E.,  Smail I.,  Best P.~N.,  Kurk J.,  Casali M.,  Ivison R.~J.,
  Coppin K.,  2008, \mn@doi [{\textbackslash}mnras]
  {10.1111/j.1365-2966.2008.13481.x}, 388, 1473

\bibitem[\protect\citeauthoryear{Gunawardhana et~al.,}{Gunawardhana
  et~al.}{2013}]{gunawardhana_galaxy_2013}
Gunawardhana M. L.~P.,  et~al., 2013, \mn@doi [{\textbackslash}mnras]
  {10.1093/mnras/stt890}, 433, 2764

\bibitem[\protect\citeauthoryear{Hopkins \& Beacom}{Hopkins \&
  Beacom}{2006}]{hopkins_normalization_2006}
Hopkins A.~M.,  Beacom J.~F.,  2006, \mn@doi [The Astrophysical Journal]
  {10.1086/506610}, 651, 142

\bibitem[\protect\citeauthoryear{Kennicutt}{Kennicutt}{1998}]{kennicutt_star_1998}
Kennicutt R.~C.,  1998, \mn@doi [Annu. Rev. Astron. Astrophys.]
  {10.1146/annurev.astro.36.1.189}, 36, 189

\bibitem[\protect\citeauthoryear{Kennicutt \& Evans}{Kennicutt \&
  Evans}{2012}]{kennicutt_star_2012}
Kennicutt R.~C.,  Evans N.~J.,  2012, \mn@doi [Annual Review of Astronomy and
  Astrophysics] {10.1146/annurev-astro-081811-125610}, 50, 531

\bibitem[\protect\citeauthoryear{Kroupa}{Kroupa}{2001}]{kroupa_variation_2001}
Kroupa P.,  2001, \mn@doi [{\textbackslash}mnras]
  {10.1046/j.1365-8711.2001.04022.x}, 322, 231

\bibitem[\protect\citeauthoryear{Leitherer et~al.,}{Leitherer
  et~al.}{1999}]{leitherer_starburst99:_1999}
Leitherer C.,  et~al., 1999, \mn@doi [{\textbackslash}apjs] {10.1086/313233},
  123, 3

\bibitem[\protect\citeauthoryear{Lilly, Le~Fevre, Hammer  \& Crampton}{Lilly
  et~al.}{1996}]{lilly_canada-france_1996}
Lilly S.~J.,  Le~Fevre O.,  Hammer F.,   Crampton D.,  1996, \mn@doi
  [{\textbackslash}apjl] {10.1086/309975}, 460, L1

\bibitem[\protect\citeauthoryear{Lonsdale~Persson \& Helou}{Lonsdale~Persson \&
  Helou}{1987}]{lonsdale_persson_origin_1987}
Lonsdale~Persson C.~J.,  Helou G.,  1987, \mn@doi [The Astrophysical Journal]
  {10.1086/165082}, 314, 513

\bibitem[\protect\citeauthoryear{Ly, Lee, Dale, Momcheva, Salim, Staudaher,
  Moore  \& Finn}{Ly et~al.}{2011}]{ly_hensuremathalpha_2011}
Ly C.,  Lee J.~C.,  Dale D.~A.,  Momcheva I.,  Salim S.,  Staudaher S.,  Moore
  C.~A.,   Finn R.,  2011, \mn@doi [{\textbackslash}apj]
  {10.1088/0004-637X/726/2/109}, 726, 109

\bibitem[\protect\citeauthoryear{Madau \& Dickinson}{Madau \&
  Dickinson}{2014}]{madau_cosmic_2014}
Madau P.,  Dickinson M.,  2014, \mn@doi [Annual Review of Astronomy and
  Astrophysics] {10.1146/annurev-astro-081811-125615}, 52, 415

\bibitem[\protect\citeauthoryear{Madau, Ferguson, Dickinson, Giavalisco,
  Steidel  \& Fruchter}{Madau et~al.}{1996}]{madau_high-redshift_1996}
Madau P.,  Ferguson H.~C.,  Dickinson M.~E.,  Giavalisco M.,  Steidel C.~C.,
  Fruchter A.,  1996, \mn@doi [{\textbackslash}mnras]
  {10.1093/mnras/283.4.1388}, 283, 1388

\bibitem[\protect\citeauthoryear{Moe \& Di~Stefano}{Moe \&
  Di~Stefano}{2017}]{moe_mind_2017}
Moe M.,  Di~Stefano R.,  2017, \mn@doi [{\textbackslash}apjs]
  {10.3847/1538-4365/aa6fb6}, 230, 15

\bibitem[\protect\citeauthoryear{Momcheva et~al.,}{Momcheva
  et~al.}{2016}]{momcheva_3d-hst_2016}
Momcheva I.~G.,  et~al., 2016, \mn@doi [ApJS] {10.3847/0067-0049/225/2/27},
  225, 27

\bibitem[\protect\citeauthoryear{Murphy et~al.,}{Murphy
  et~al.}{2011}]{murphy_calibrating_2011}
Murphy E.~J.,  et~al., 2011, \mn@doi [The Astrophysical Journal]
  {10.1088/0004-637X/737/2/67}, 737, 67

\bibitem[\protect\citeauthoryear{Osterbrock \& Ferland}{Osterbrock \&
  Ferland}{2006}]{osterbrock_astrophysics_2006}
Osterbrock D.~E.,  Ferland G.~J.,  2006, Astrophysics of gaseous nebulae and
  active galactic nuclei

\bibitem[\protect\citeauthoryear{Panter, Jimenez, Heavens  \& Charlot}{Panter
  et~al.}{2007}]{panter_star_2007}
Panter B.,  Jimenez R.,  Heavens A.~F.,   Charlot S.,  2007, \mn@doi
  [{\textbackslash}mnras] {10.1111/j.1365-2966.2007.11909.x}, 378, 1550

\bibitem[\protect\citeauthoryear{Salim et~al.,}{Salim
  et~al.}{2007}]{salim_uv_2007}
Salim S.,  et~al., 2007, \mn@doi [The Astrophysical Journal Supplement Series]
  {10.1086/519218}, 173, 267

\bibitem[\protect\citeauthoryear{Skelton et~al.,}{Skelton
  et~al.}{2014}]{skelton_3d-hst_2014}
Skelton R.~E.,  et~al., 2014, \mn@doi [ApJS] {10.1088/0067-0049/214/2/24}, 214,
  24

\bibitem[\protect\citeauthoryear{Sobral, Smail, Best, Geach, Matsuda, Stott,
  Cirasuolo  \& Kurk}{Sobral et~al.}{2013}]{sobral_large_2013}
Sobral D.,  Smail I.,  Best P.~N.,  Geach J.~E.,  Matsuda Y.,  Stott J.~P.,
  Cirasuolo M.,   Kurk J.,  2013, \mn@doi [Monthly Notices of the Royal
  Astronomical Society] {10.1093/mnras/sts096}, 428, 1128

\bibitem[\protect\citeauthoryear{Stanway \& Eldridge}{Stanway \&
  Eldridge}{2018}]{stanway_re-evaluating_2018}
Stanway E.~R.,  Eldridge J.~J.,  2018, \mn@doi [{\textbackslash}mnras]
  {10.1093/mnras/sty1353}, 479, 75

\bibitem[\protect\citeauthoryear{Stanway, Eldridge, Greis, Davies, Wilkins  \&
  Bremer}{Stanway et~al.}{2014}]{stanway_interpreting_2014}
Stanway E.~R.,  Eldridge J.~J.,  Greis S. M.~L.,  Davies L. J.~M.,  Wilkins
  S.~M.,   Bremer M.~N.,  2014, \mn@doi [{\textbackslash}mnras]
  {10.1093/mnras/stu1682}, 444, 3466

\bibitem[\protect\citeauthoryear{Tojeiro, Heavens, Jimenez  \& Panter}{Tojeiro
  et~al.}{2007}]{tojeiro_recovering_2007}
Tojeiro R.,  Heavens A.~F.,  Jimenez R.,   Panter B.,  2007, \mn@doi
  [{\textbackslash}mnras] {10.1111/j.1365-2966.2007.12323.x}, 381, 1252

\bibitem[\protect\citeauthoryear{Wilkins, Trentham  \& Hopkins}{Wilkins
  et~al.}{2008a}]{wilkins_evolution_2008}
Wilkins S.~M.,  Trentham N.,   Hopkins A.~M.,  2008a, \mn@doi
  [{\textbackslash}mnras] {10.1111/j.1365-2966.2008.12885.x}, 385, 687

\bibitem[\protect\citeauthoryear{Wilkins, Hopkins, Trentham  \&
  Tojeiro}{Wilkins et~al.}{2008b}]{wilkins_extragalactic_2008}
Wilkins S.~M.,  Hopkins A.~M.,  Trentham N.,   Tojeiro R.,  2008b, \mn@doi
  [{\textbackslash}mnras] {10.1111/j.1365-2966.2008.13890.x}, 391, 363

\bibitem[\protect\citeauthoryear{Wright, Driver  \& Robotham}{Wright
  et~al.}{2018}]{wright_gama/g10-cosmos/3d-hst:_2018}
Wright A.~H.,  Driver S.~P.,   Robotham A. S.~G.,  2018, \mn@doi [Monthly
  Notices of the Royal Astronomical Society] {10.1093/mnras/sty2136}, 480, 3491

\bibitem[\protect\citeauthoryear{Yu \& Wang}{Yu \&
  Wang}{2016}]{yu_inconsistency_2016}
Yu H.,  Wang F.~Y.,  2016, \mn@doi [ApJ] {10.3847/0004-637X/820/2/114}, 820,
  114

\makeatother
\end{thebibliography}

\end{document}